\documentclass[manuscript]{aastex}

\newcommand{\logg}{{\rm log}~g}
\newcommand{\teff}{T_{\rm eff}}

\shorttitle{Superflares on Solar Type Stars Observed with Kepler}
\shortauthors{Shibayama et al.}

\usepackage{ulem}

\begin{document}

\title{Superflares on Solar Type Stars Observed with Kepler \\
I. Statistical Properties of Superflares}

\author{Takuya Shibayama\altaffilmark{1}, Hiroyuki Maehara\altaffilmark{2,3},
Shota Notsu\altaffilmark{1}, Yuta Notsu\altaffilmark{1}, Takashi Nagao\altaffilmark{1}, 
Satoshi Honda\altaffilmark{2,4}, Takako T. Ishii\altaffilmark{2}, Daisaku Nogami\altaffilmark{2}, Kazunari Shibata\altaffilmark{2}}
\altaffiltext{1}{Department of Astronomy, Kyoto University, Sakyo, Kyoto, 606-8502, Japan}
\altaffiltext{2}{Kwasan and Hida Observatory, Kyoto University, Yamashina, Kyoto, 607-8471, Japan}
\altaffiltext{3}{Kiso Observatory, Institute of Astronomy, School of Science, The University of Tokyo, 10762-30, Mitake, Kiso-machi, Kiso-gun, Nagano 397-0101, Japan}
\altaffiltext{4}{Center for Astronomy, University of Hyogo, 407-2, Nishigaichi, Sayo-cho, Sayo, Hyogo, 679-5313, Japan}

\email{shibayama@kwasan.kyoto-u.ac.jp}

\begin{abstract}

By extending our previous study by \citet{maehara2012}, we searched for
superflares on G-type dwarfs (solar type stars) using Kepler data for a longer period
(500 days) than that (120 days) in our previous study.  
As a result, we found 1547 superflares on 279 G-type dwarfs, 
which are much more than previous 365 superflares on 148 stars. 
Using these new data, we studied the statistical properties of
occurrence frequency of superflares, and basically confirmed 
the previous results, i.e., the occurrence frequency ($dN/dE$) 
of superflares vs flare energy ($E$) shows power-law distribution with 
$dN/dE \propto E^{-\alpha}$, where $\alpha \sim 2$. 
It is interesting that this distribution is roughly on the
same line as that for solar flares.  
In the case of the Sun-like stars (with 
surface temperature 5600-6000K and slowly rotating with a period
longer than 10 days), the occurrence frequency of superflares with 
energy of $10^{34}-10^{35}$ erg is once in 800-5000 years.
 We also studied long term (500 days) stellar brightness variation 
of these superflare stars, and found that in some G-type dwarfs 
the occurrence frequency of superflares was extremely high, 
$\sim$ 57 superflares in 500 days (i.e., once in 10 days).
In the case of Sun-like stars, the most active stars show
the frequency of one superflares (with $10^{34}$ erg) 
in 100 days. There is evidence that these superflares have
extremely large starspots with a size about 10 times 
larger than that of the largest sunspot.  
We argue that the physical origin of extremely high
occurrence frequency of superflares in these stars may be
attributed to the existence of extremely large starspots.

\end{abstract}


\section{Introduction}

Flares are explosions on the stellar surface with intense release of 
the magnetic energy stored near starspots in the outer atmosphere of stars
\citep[e.g.,][]{shibata2002, gershberg2005, benz2010, shibata2011}. 
During these flare events, various observable effects occur over 
a wide range in wavelength.
On the Sun,  our ability to spatially resolve the surface allows 
flares to be observed in rich detail with high cadence, 
and so much of our understanding of flare processes 
and their consequences come from observations of solar flares. 

The typical total energy of solar flare ranges from $10^{29}$ to $10^{32}$ erg
\citep[e.g.,][]{shibata2002},
and the duration is several minutes to several hours. 
There have been no observation of solar flares more energetic than $10^{32}$ erg. 
On the other hand, it is known that such more energetic flares occur on a variety of stars \citep{schaefer1989,landini1986}.
\citet{schaefer2000} reported 9 flare candidates on G-type dwarfs
, i.e., solar type stars (5100 K $\leq \teff <$ 6000 K and $\logg >$ 4.0),
with a total energy of $10^{33}$ to $10^{38}$ erg and, called superflares. 
The data source of their study was, however, various and ambiguous 
(e.g. photography,  X-ray,  visual and so on) and the number of discovered flares 
is small so that statistical properties, such as the frequency 
and energy distribution of superflares, are not well known.

Since solar flares lead to magnetic storm on the Earth, 
extensive damage may be caused in our civilization by large solar flares
\citep[e.g.,][]{baker2004}. 
Geomagnetic storms induce electrical currents that can have significant impact 
on electrical transmission equipments,  leading to a wide spread blackout. 
On March 13, 1989, Quebec in Canada, 6 million people were without electric power for 9 hours 
as a result of a huge geomagnetic storm \citep{allen1989}. 
Moreover, in 1859, our civilization experienced the largest flare on the Sun ($\sim10^{32}$ erg ), 
called Carrington flare after the discoverer \citep{carrington1859},
which caused the largest magnetic storm in recent 200 years 
\citep{tsurutani2003}, leading to failure in telegraph system all over Europe
and North America \citep{loomis1861}. 

Even solar flares of up to $10^{32}$ erg have strong effects on the Earth, 
like this.  If  superflares would occur on the Sun, 
our civilization would suffer from much more severe damages. 
Hence it is very important to study the statistical properties of superflares
on G-type dwarfs, especially to reveal whether superflares would
really occur on Sun-like stars 
which are defined as the stars with
surface temperature, 5600 K $\leq \teff <$ 6000 K,
surface gravity, $\logg >$ 4.0,
and rotational period, P $>$ 10 d. 
Recently, using Kepler data, 
\citet{maehara2012} discovered 365 superflares ($10^{33}-10^{36}$ erg)  
on 148 G-type dwarfs, which
are much more than 9 superflares on G-type dwarfs discovered by \citet{schaefer2000}.
Among them, 14 superflares occurred on 10 Sun-like stars.
Hence \citet{maehara2012} successfully analyzed the statistical properties of
superflares on G-type dwarfs for the first time, revealing that

(1) The occurrence frequency of these superflares as  a function of their total energy 
is quite similar to that of solar flares ($dN/dE \propto E^{-\alpha}$, $\alpha \sim 2$). 

(2) These superflare stars show quasi periodic brightness variation, which may be evidence of
very big starspots with stellar rotation.  

(3) It is found that superflares occur on Sun-like stars  with frequency such that superflares with energy $10^{34}$-$10^{35}$ erg
 (100-1000 times of the largest solar flare) occur once in 800-5000 years. 

(4) There is no hot Jupiter around these superflare stars, suggesting that hot Jupiters
are rare in superflare stars. 

The result (4)  is important because \citet{rubenstein2000} suggested that
G-type dwarfs with hot Jupiter companion are good candidates of superflare stars.
With this idea, they proposed that the Sun would never produce superflares
since our Sun does not have hot Jupiter. The result (4) suggests that their argument
is not necessarily correct. 

In spite of these new observations, \citet{schaefer2012} still argued that it was unlikely
that superflares would occur on our present Sun.
On the other hand,  \citet{shibata2013}
concluded that we cannot reject the possibility that superflares of 
$10^{34} - 10^{35}$ erg  
would also occur on our present Sun with  frequency once in a few 1000 years
in view of the present theory of flares and dynamo. 

The new observations of superflares by \citet{maehara2012} opened the renewed 
interest in the possibility of superflares on our Sun as well as on G-type dwarfs.
The latter subject is important from viewpoint of the habitability of exoplanets 
around G-type dwarfs. 

The purpose of this paper is to study statistical properties of superflares on
G-type dwarfs and Sun-like stars in more detail than before, 
extending previous study by \citet{maehara2012} (based on 120 days data) 
using longer term (500 days)  observational data in Kepler mission.
We found 1547 superflares on 279 G-type dwarfs, and 
44 superflares on 19 Sun-like stars
, which are much more than those discovered by 
\citet{maehara2012} who discovered 365 superflares on 148 G-type dwarfs 
and 14 superflares on 10 Sun-like stars. 

In section 2, we describe method of analysis, and in section 3,  we present
typical light curves of superflares and main results of statistical 
analysis on occurrence frequency.  We also examined long term
brightness variation and superflares 
on the most active G-type dwarfs and Sun-like stars, and found remarkable
results that superflares occur once in 10 - 100 days in these active stars.
In section 4, we discuss physical meaning of these results and compared 
these results with time variation of solar flare occurrence frequency 
as a function of solar activity level. 
Finally, in Appendix, we show short time cadence Kepler data for
two typical superflares, and confirmed that long time cadence
data basically grasp the basic properties of superflares.

\section{Method}

\subsection{Kepler Data}
The Kepler carries an optical telescope with 95cm aperture and 
105 square deg field-of-view (about 12 deg diameter), 
which is in the Cygnus, Lyra and Draco. This space craft was launched by NASA 
in Spring 2009 to search for exoplanets by finding planetary transit events, 
a faint decrease in brightness due to a crossing planet. 
Since the orbiting planet is very small compared with the host star, 
the luminosity decrease by planetary transit is usually less than 
one hundredth of the total brightness of the star. In addition, 
the planet passes in front of the star only when 
the orbit is nearly parallel with the line of sight.
Kepler is therefore designed to obtain high-precision 
and long-period light curves of many stars. 
The typical precision is 0.1 mmag for a star of 12 mag and 
the number of observed stars is more than 160,000 \citep{koch2010}.  
The time resolution is about 30 min and 1 min. 
The resultant light curves are useful in detecting not only planetary transits 
but also other small brightness variations like stellar flares.
In fact, there are some previous studies about stellar flare using Kepler data 
\citep{walkowicz2011, balona2012}, which focus on flares on early or late type stars.

\subsection{Analysis Method}

The data we used were taken during the period from 
May 2009 to September 2010. 
We retrieved the data from the Multimission Archive at Space Telescope 
Science Institute (MAST) and 
analyzed the long-cadence (the time resolution is about 30 min) corrected flux of 
9511, 75598, 82811, 82586, 89188, 86248, 82052 stars 
in quarters 0, 1, 2, 3, 4, 5, and 6 respectively 
(all public light curves of G-type dwarfs observed by Kepler). 
The number of G-type dwarfs and monitoring period of each quarter are shown in Table \ref{tab:nstar}.
In Table \ref{tab:nstar}, we also show the beginning and the end dates of each quarter, which 
are shown in the Kepler Data Release Notes \footnote{http://archive.stsci.edu/kepler/data\_release.html}.  

Since our aim of this study is to detect superflares on G-type dwarfs, 
we selected G-type dwarfs in all the observed stars using 
Kepler Input Catalog \citep{brown2011}. 
The condition is 5100 K $\leq \teff <$ 6000 K and $\logg >$ 4.0 and 
the number of selected stars is about 80,000 in 160,000.

There are various flare detection algorithms which detect flares 
from light curves \citep{osten2012,walkowicz2011}.
In these algorithms, light curves are analyzed after detrended and
they searched for the case where the relative flux become statistically 
larger than a certain threshold for two or more times consecutively.
The time resolution of the light curve of the Kepler is about 30 min and some stars exhibit 
short periodic brightness variation (several hours).
This period is comparable with the time scale of flare duration so that   
detrending light curve has a possibility to detrend the superflares themselves.
To avoid misdetection of short stellar brightness variation and not to overlook large flares,
we calculated the distributions of brightness variation between all pairs of consecutive data points 
after creating light curves of all selected stars using the corrected flux in the data. 
The threshold of the flare was determined to be three times the value at the top 1\% of the distribution.
This threshold was chosen as a result of test run so as not to misdetect other brightness variation. 
We also calculated the standard deviation of the distribution ($\sigma _{\rm diff}$)
for defining the end time of a flare.

\begin{figure}
  \plotone{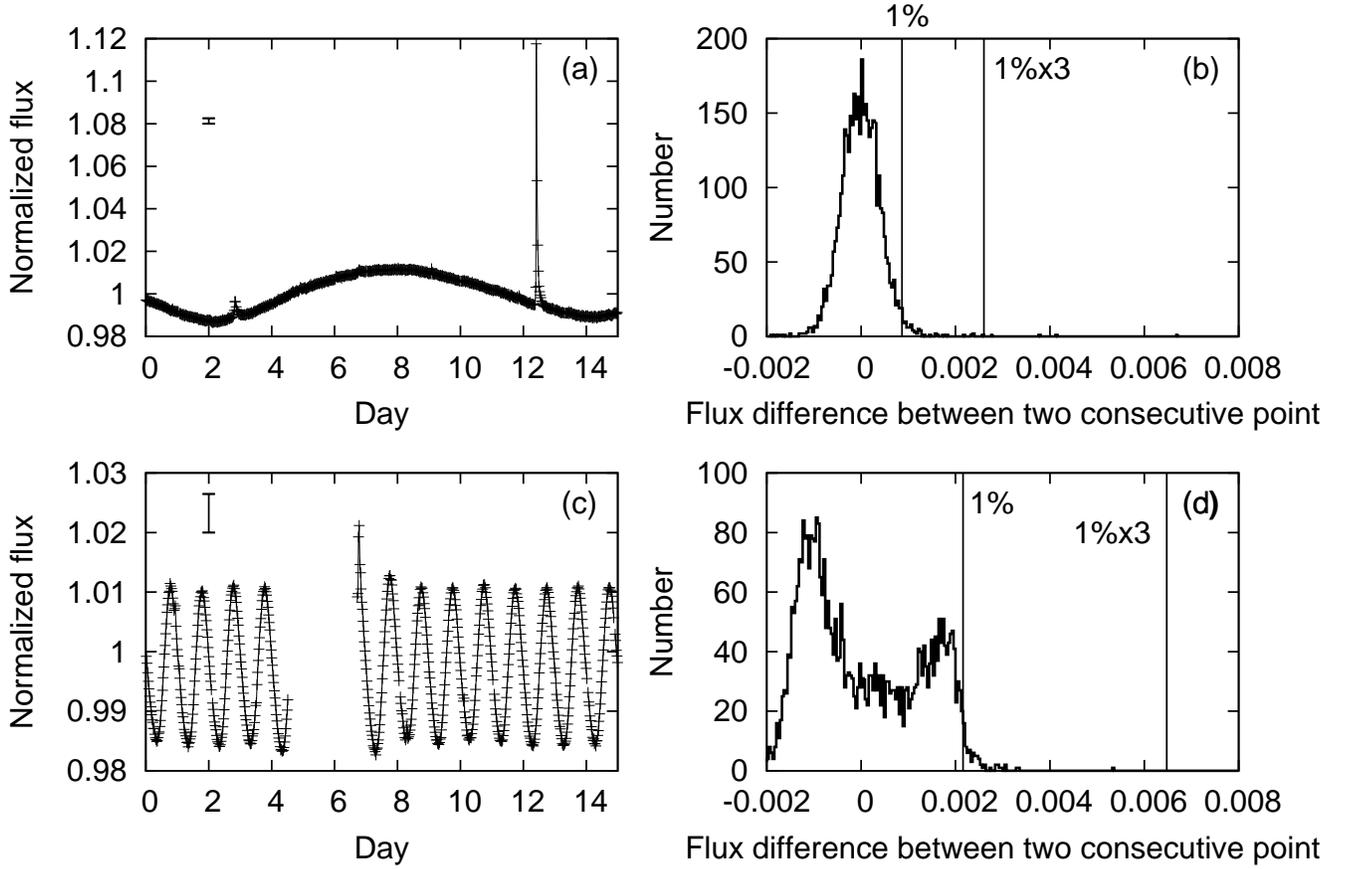}
  \caption{Explanatory figures of our flare-search method. 
The histograms in the right panels show the number distribution of brightness variation
between all pairs of 2 consecutive data points. 
We set the threshold as three times of the value at the top 1\% of the distribution, and
brightenings larger than the threshold are detected as flare candidates.
Bars at $t=2$ of the left panels correspond to the threshold determined by the right panels.
(a) Light curve of KIC9603367. (b) Histogram of the number distribution of the brightness difference between all pairs of 2 consecutive data points in Figure (a).
(c) Another example of light curve of KIC 4830001. 
(d) The same as Figure (b), but for data in Figure (c).
Since the ordinary flux difference is much larger than in Figure (b), 
only the flares with a much larger rising rate are detected.
Figure (d) shows a distribution with two peaks, which correspond to the rates of typical 
increase and decrease in the brightness.
}
  \label{fig:method}
\end{figure}

Figure \ref{fig:method} shows a schematic figure of our method.
Figure \ref{fig:method} (a) and (b) show a light curve of KIC 9603367 and
the distribution of brightness variations between consecutive data points.
Figure \ref{fig:method} (c) and (d) are the same as (a) and (b) but for
KIC 4830001.
We show examples of determination of threshold in figures.
Vertical lines in the right panels written ``1\%'' correspond to the value of top 1\%,
and the other vertical lines written ``1\%x3'' show threshold of flare detection.
Bars in the left panels at $t=2 {\rm d}$ show the threshold decided by the right panels.
Although the amplitude of brightness variation of KIC 9603367 and KIC 4830001
are almost the same, the period of the brightness variation of KIC 9603367
(12.2 d) is longer than that of KIC 4830001 (1.0 d).
Since the brightness variation between two consecutive observations
in the stars with short-period variations (e.g. KIC 4830001) are 
larger than that of the stars with long-period variations
(e.g. KIC 9603367), the distribution of
brightness variations in short-period systems tend to spread wider than that of
long-period systems, and the detection threshold of flares in 
short-period systems, therefore, becomes larger than that of flares in long-period systems.
The detection threshold also depends on the amplitude of brightness variations.
The thresholds in the stars showing large-amplitude variations tend to be 
larger than those in the small-amplitude systems.

The detection limit of this algorithm is determined from
the value of the threshold divided by the average flux of the light curve.
This detection limits are between 0.001 to 0.01 for most of our targets, 
therefore, flares are detected almost completely 
when the amplitudes are more than 1\% of the average flux.
The typical total energy of flare of 1\%-amplitude is $5 \times 10^{34}$ erg. 

The time at which the flux first exceeds the threshold is determined as the flare start time, 
and we removed long term brightness variations around 
the flare by fitting with spline curves of three points.
Here three points are determined by the average of 
some data points just before the beginning of the flare and 
around 5 hours and 8 hours after the peak of the flare.
After detrended, the flare end time was defined to be 
the time at which the flux residuals become smaller than $3~\sigma _{\rm diff}$.
We analyzed only flares with the duration longer than 0.05 days
(at least 2 consecutive data points exceed the threshold) and
excluded flare candidates which have shorter decline phase than the increase phase.

The pixel scale of the Kepler CCDs is about 4 arc-seconds and the typical
photometric aperture for a 12-mag star contains about 30 pixels \citep{van cleve2009}.
This suggests that the brightness variation of neighboring stars can affect 
the flux of the target star.
If there is a flare-star near the target and if large-amplitude flares occur
on the flare-star, Kepler can detect fake flares on the target star.
In order to avoid these false events, we chose stars without neighboring
stars within 12 arc-second for the analysis. The total number of samples
which satisfy this condition was about 30 \%.
The spatial distance between a pair of stars is calculated from the star properties in KIC.
Superflares are rare event, therefor, the probability is very low 
that stars visually nearby exhibit superflare at the same 30 min cadence.
Such flare candidates are thought to be misdetected because of instrumental characteristics or contaminations of other sources.
Although the reason of this instrumental characteristics are uncertain, 
light curves occasionally show discontinuity, which is thought to be 
caused by the variation of pointing accuracy or cosmic ray.
We therefore excluded pairs of flares which occurred at the same time
and whose spatial distance is less than 24 arc-seconds.

After removing the candidates of flares on neighboring stars,
we checked light curves of flare candidates by eye, and
examined the pixel level data of each G-type dwarf exhibiting flares by eye.
Some contaminations of eclipsing binaries or transit events are found from light curves.
If spatial distribution of brightness on the CCDs is different between at flares and at the quiescence,
the flare is revealed to be a brightness variation of another source.
A few percents and about 10 \% of flare candidates are removed by
checking light curves and pixel level data respectively.

We calculated periods of brightness variation from light curves, which correspond to the rotational period 
in the case of single star with dark spot. 
The variation period was calculated by the discrete Fourier transform (DFT) method, 
and we chose the largest peak as the rotational period of the star 
when the peak is enough larger than the error level.

\begin{figure}
  \plotone{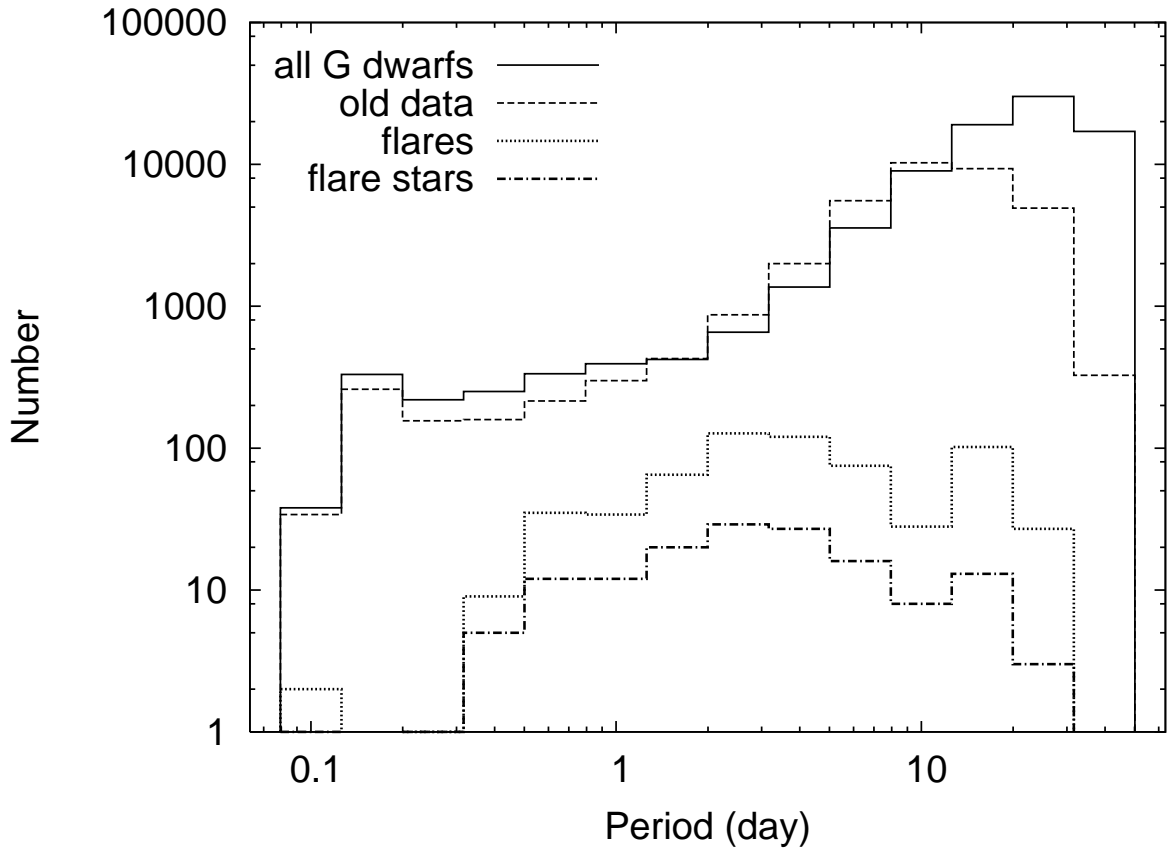}
  \caption{
      Number distributions of all G-type dwarfs (solid line),
      flare stars (dash-dotted line) and flares (dotted line) 
      observed by Kepler as a function of rotational period.
      The periods of these stars are calculated from new data. 
      We selected flares whose total energy are larger 
      than $5\times 10^{34}$ erg.
      Dashed line indicates the distribution of period calculated 
      from the old data, which is calibrated by the pipeline of the previous 
      version. 
  }
  \label{fig:period}
\end{figure}

Figure \ref{fig:period} shows 
number distributions of all G-type dwarfs (solid line), 
flare stars (dash-dotted line) and flares (dotted line) 
observed by Kepler as a function of rotational period. 
The periods of these stars are calculated from new data. 
Dashed line indicates the distribution of period calculated 
from the old data, which is calibrated by the pipeline of 
the previous version. 
In this figure, we selected only flares whose total energy are more than
$5\times 10^{34}$ erg.
We obtained public data of release 14 and 16 in this study. Dashed  
line shows distribution of period calculated from light 
curves of release from 4 to 9, which were used in Maehara 
et al. (2012). In the release from 4 to 9, the data are 
calibrated by old pipeline, and long term brightness variation 
longer than 15 days tends to be removed (van Cleve et al. 2010). 
Hence, the number of stars with period more than 20 days was 
small (Table \ref{tab:period}, Figure \ref{fig:period}). 
The reduction algorithm for the new Kepler 
data in release 14 and 16 is improved (Jenkins et al. 2012; 
Stumpe et al.2012). As a result, long term brightness variation 
has been detected much better than before, and more stars 
with period longer than 20 days are, therefore, 
detected from new data (release 14 and 16) than old data
(release 4 to 9).
We chose 10 days as the threshold of Sun-like stars.
Rotational period of 10 days corresponds to a radial velocity of $\sim5$km/s, 
and this threshold is enough for discrimination of slowly rotating stars from 
rapidly rotating stars (e.g. $v > 10$km/s).
This threshold is consistent with the threshold of ``ordinary G-type dwarfs'' in \citet{schaefer2000}.

\subsection{Energy Estimation}
We estimated the total energy of each flare from stellar luminosity, amplitude and duration of flares
by assuming that the spectrum of white-light flares can be described by black body radiation 
with an effective temperature of 10,000 K ($T_{\rm flare}$) \citep{mochnacki1980,hawley1992}.
Assuming the case of the effective temperature is 9,000 K, 
the estimated energy becomes 66\% of our result.
This is the possible systematic uncertainty of our energy estimate.

Assuming that the star is a blackbody radiator,
the bolometric flare luminosity ($L_{\rm flare}$) is calculated from 
$T_{\rm flare}$ and the area of flare ($A_{\rm flare}$) as the following equation,
\begin{equation}
L_{\rm flare} = \sigma_{\rm SB} T_{\rm flare}^4 A_{\rm flare}~,
\end{equation}
where $\sigma_{SB}$ is the Stefan-Boltzmann constant.
Since the star is not the blackbody radiator,
this estimate is not accurate and may have an error of a few tens of percent.
For the estimate of $A_{\rm flare}$ we use observed\footnotemark[1] luminosity of star ($L'_{\rm star}$)
, flare ($L'_{\rm flare}$) and flare amplitude of 
\footnotetext[1]{These observed luminosities have instrumental convolution factor.}
the light curve of Kepler after detrended ($C'_{\rm flare}$).
These are calculated from these equations, 
\begin{eqnarray}
L'_{\rm star} &=& \int R_{\lambda}B_{\lambda(T_{\rm eff})}d\lambda \cdot \pi R_{\rm star}^2~, \\
L'_{\rm flare} &=& \int R_{\lambda}B_{\lambda(T_{\rm flare})}d\lambda \cdot A_{\rm flare}~, and\\
C'_{\rm flare} &=& L'_{\rm flare}/L'_{\rm star}~, 
\end{eqnarray}
where $\lambda$ is the wavelength, $B_{\lambda(T)}$ is the Plank function, and
$R_{\lambda}$ is the response function of the Kepler instrument.
The instrument has a spectral bandpass from 400 nm to 850 nm \citep{van cleve2009}.
We can estimate $A_{\rm flare}$ from these equation as follows,
\begin{equation}
A_{\rm flare} = C'_{\rm flare} \pi R^2 \frac{\int R_{\lambda}B_{\lambda(T_{\rm eff})}d\lambda}{\int R_{\lambda}B_{\lambda(T_{\rm flare})}d\lambda}~.
\end{equation}

$L_{\rm flare}$ can be estimated from equations (1) and (5), 
and $C'_{\rm flare}$ is a function of time, therefore, 
$L_{\rm flare}$ is also a function of time.
Total bolometric energy of superflare ($E_{\rm flare}$) 
is an integral of $L_{\rm flare}$ during the flare duration,
\begin{equation}
E_{\rm flare} = \int_{\rm flare} L_{\rm flare}(t) dt~.
\end{equation}

According to \citet{brown2011}, $1-\sigma$ uncertainties of the stellar radius and the surface temperature 
in the Kepler Input Catalog are about $\pm$ 0.2 dex and $\pm$ 200 K respectively. 
Hence the uncertainty in the stellar luminosity can be estimated to be about $\pm$ 60 \%. 
Although the time-cadence of the Kepler data was not sufficiently short 
to determine the peak and duration of flares, 
the uncertainties of the amplitude and duration of flares are much smaller than 
the uncertainty in the stellar luminosity because, when we calculate the flare energy, 
we integrate the brightness variation due to the flare over time from the start to end of the flare. 
Therefore total uncertainty in the flare energy is about $\pm 60$ \% and the occurrence frequency of superflares 
in unit energy range includes not only the counting error 
(the square root of the number of flares) but also the uncertainty of the flare energy.

\section{Results}
\subsection{Typical Example}

We detected 1547 superflares on 279 G-type dwarfs, 
including 44 superflares on 19 Sun-like stars
from the Kepler long cadence data.
The monitoring duration was about 500 days (quarters 0 to 6).
365 superflares were detected from 148 stars in a monitoring duration of about 120 days by \citet{maehara2012}.
The number of superflares in this study is more than four times of that of our previous study \citep{maehara2012},
while the number of flare-generating stars is less than twice.
The Kepler is monitoring the same stars and some active stars exhibit more than a few flares. 
101 superflare stars show only one superflare, while the number of superflare stars that exhibit more than one superflare is 178, 
and 8 Sun-like stars show more than one superflare in 19 Sun-like superflare stars.
All flares are listed in the online table with parameters of stars and flares. 
BV Amp in the online table indicates the amplitude of stellar brightness variation
in the quarter when the flare is detected. 
This value basically corresponds to the apparent fraction of 
the spot area to the that of the disk. 
The definition of this value is as follows.
We calculated $\Delta F(t)~(= F(t) - F_{\rm av})$, where $F(t)$ is flux of light (stellar brightness) at date $t$ and $F_{\rm av}$ is average flux. 
Then, excluding data points whose absolute values of $\Delta F$ are included in top 1\% of them,
we calculated $(\max(\Delta F) - \min(\Delta F))/F_{\rm av}$, which is defined as 
BV Amp of the star in the quarter.

\begin{figure}
  \begin{center}
  \epsscale{0.7}
    \plotone{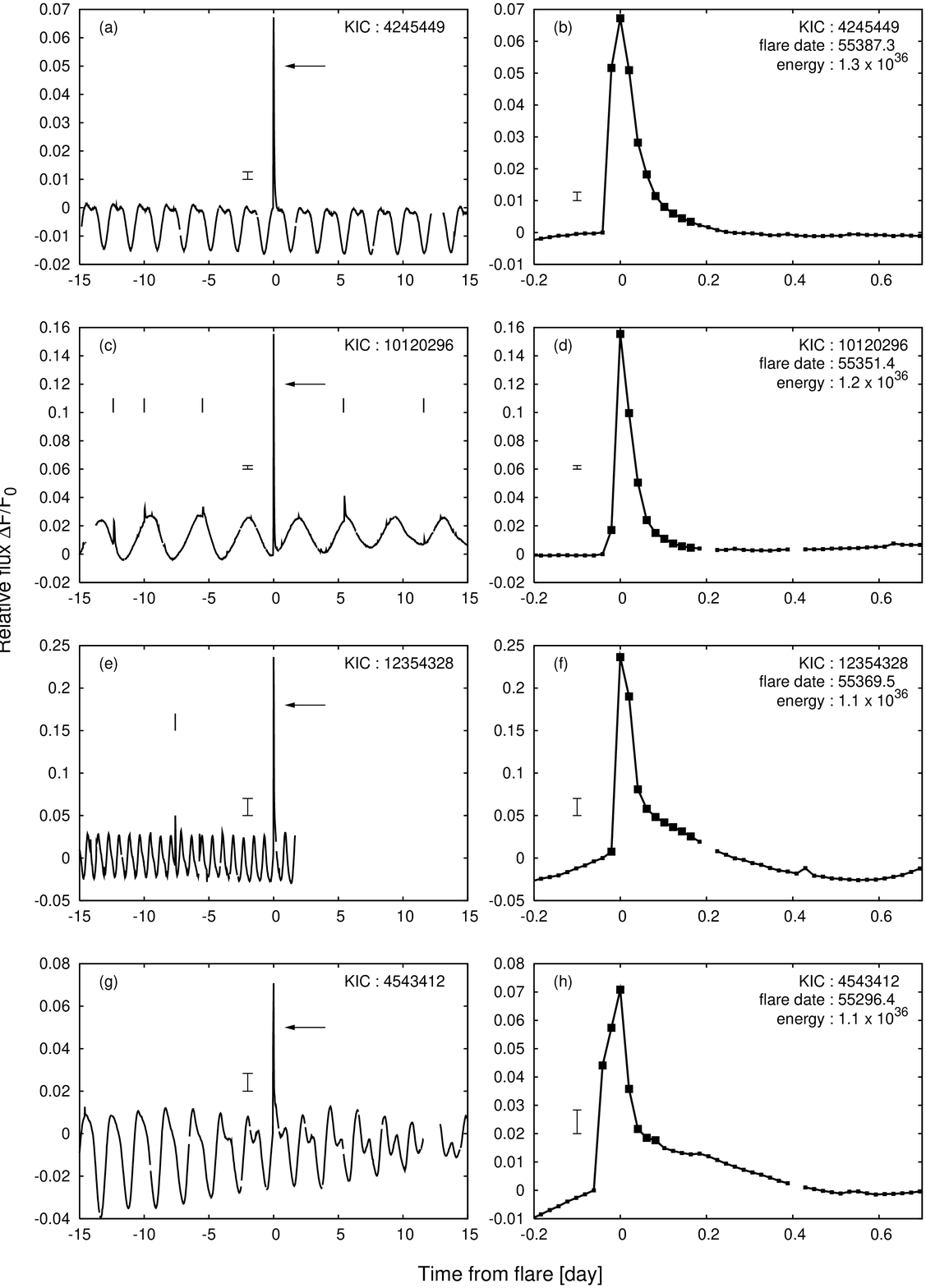}
  \end{center}
  \caption{Light curves of most energetic superflares.
  Horizontal and vertical axes correspond to days from the flare, and
  stellar brightness normalized by the brightness just before the flare ($F_0$) respectively. 
  Panels in the left side show the 30 days time variation of
  stellar brightness, while panels in the right side show detailed brightness
  variation of a flare in a short period (0.9 days).
  Vertical short lines in some left panels show the time at which flare detected.
  Each bar with heads locate just before the flare shows the detection threshold 
  of superflare for the light curve.
  The star ID (Kepler ID), the Julian date of flare peak and released total flare energy are 
  shown in the upper right corner of figures in the right column.}
  \label{fig:lf}
\end{figure}

The numbers of detected superflares
and stars exhibit these flares 
are listed in Table \ref{tab:superflare}. 
The fractions of superflare stars among all Sun-like stars, all
G-type dwarfs, etc. are shown in Table \ref{tab:allstars}
with numbers of flare stars and all observed stars in the category.
In Figures \ref{tab:superflare} and \ref{tab:allstars},
G-type dwarfs and Sun-like stars correspond to 
the lower right and middle left cells respectively.
Figure \ref{fig:lf} shows four 
light curves of most energetic superflares
with Kepler ID, flare peak date and total flare energy in each right panel
(see Table \ref{tab:starparameter} for detailed parameters of these stars.).
Left panels indicate long period light curves (30 days) and 
right panels show enlarged light curves of superflares during about a day.
Vertical short lines in some left panels in Figure \ref{fig:lf} show the time at which 
flare is detected with our detection algorithm.
Bars at $t=2$ of left panels and $t=0.1$ of right panels show flare detection thresholds 
of the light curves. 

\begin{figure}
  \begin{center}
  \epsscale{0.8}
    \plotone{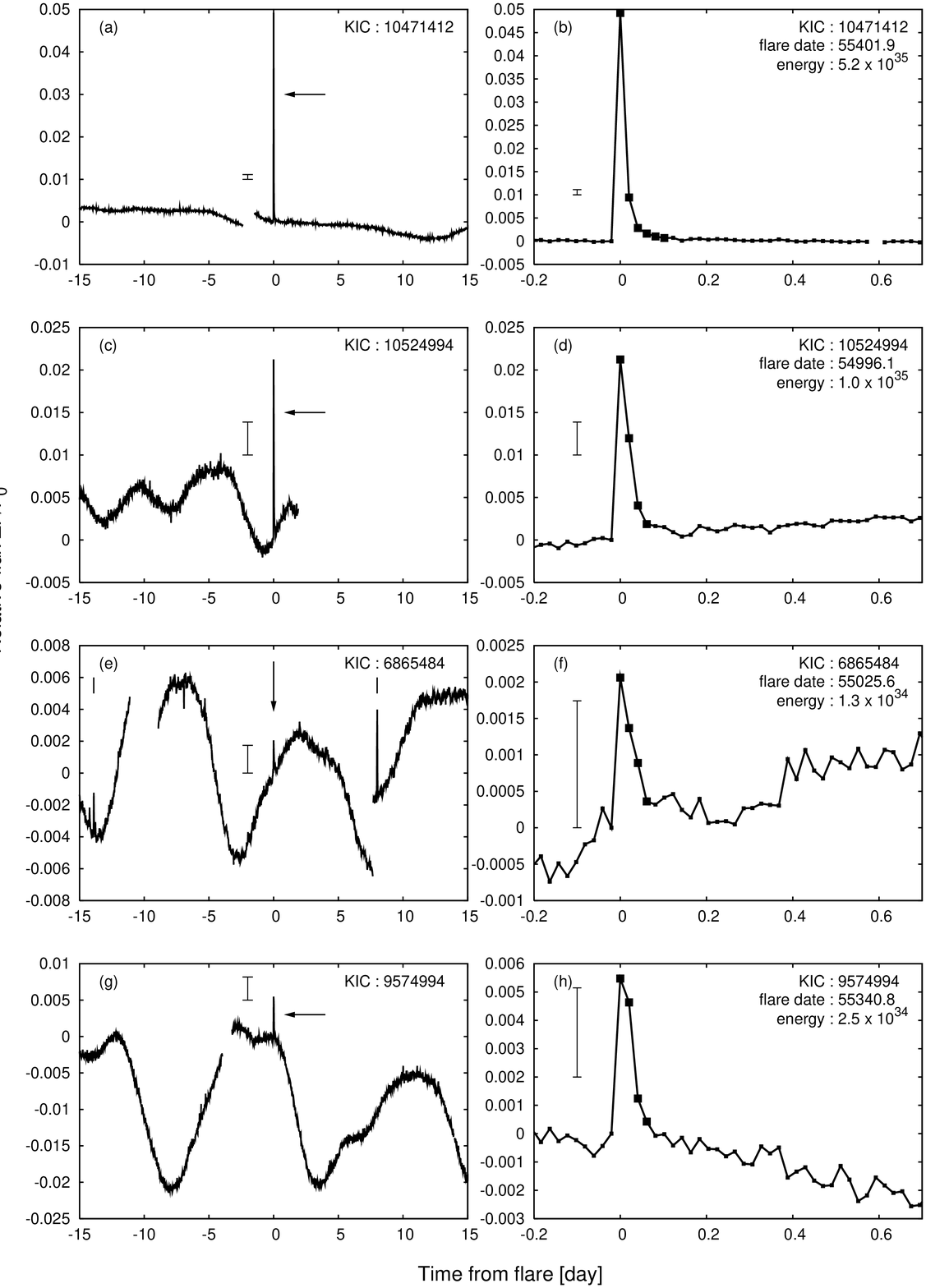}
  \end{center}
  \caption{Same as Figure \ref{fig:lf} but for largest superflares on Sun-like stars
(i.e., stars with surface temperature 5600-6000 K and the rotational 
period longer than 10 days).
Upper 4 panels show light curves of most energetic superflare on Sun-like stars, 
while lower panels show superflares on Sun-like stars with longer brightness variation period. 
}
  \label{fig:sllf}
\end{figure}

Various different periodic brightness variations are observed on many flare stars.
These variations suggest the existence of large starspots on superflare stars \citep{rodono1986}.
There are, however, other mechanisms causing the brightness variation such as 
orbital motion of a binary system, eclipse by an accompanying star \citep{kopal1959} or stellar pulsation.
Here stellar pulsation can be excluded because the pulsation period of 
the G-type dwarfs is shorter than a few hours \citep{unno1989}. 
The possibility of the brightness variation being due to rotation 
must be carefully distinguished from its being due to orbital motion, 
on the basis of the difference in the shape of the light curves \citep{debosscher2011}. 
Figure \ref{fig:sllf} shows examples of superflares on Sun-like stars. 
The durations of the detected superflares are typically a few hours, 
and their amplitudes are generally of order 0.1 $\sim$ 1 \% of the stellar luminosity. 

A major concern is the contamination from unresolved sources such as a low-mass companion.
We discussed this topic in the supplementary information of \citet{maehara2012}, and
concluded that most of the flares we detected are thought to be produced by G-type dwarfs 
because of the positive correlation between the amplitude of stellar brightness modulations and the flare frequency
in Figure 5b of \citet{maehara2012}'s Supplementary Information.
Furthermore, our results show dependence of flare occurrence frequency 
on the effective temperature of the G-dwarfs (See Table \ref{tab:allstars}).
This dependence cannot be explained by the contamination by low-mass companion.

\subsection{Occurrence Frequency of Superflares}

\begin{figure}
\epsscale{0.7}
\plotone{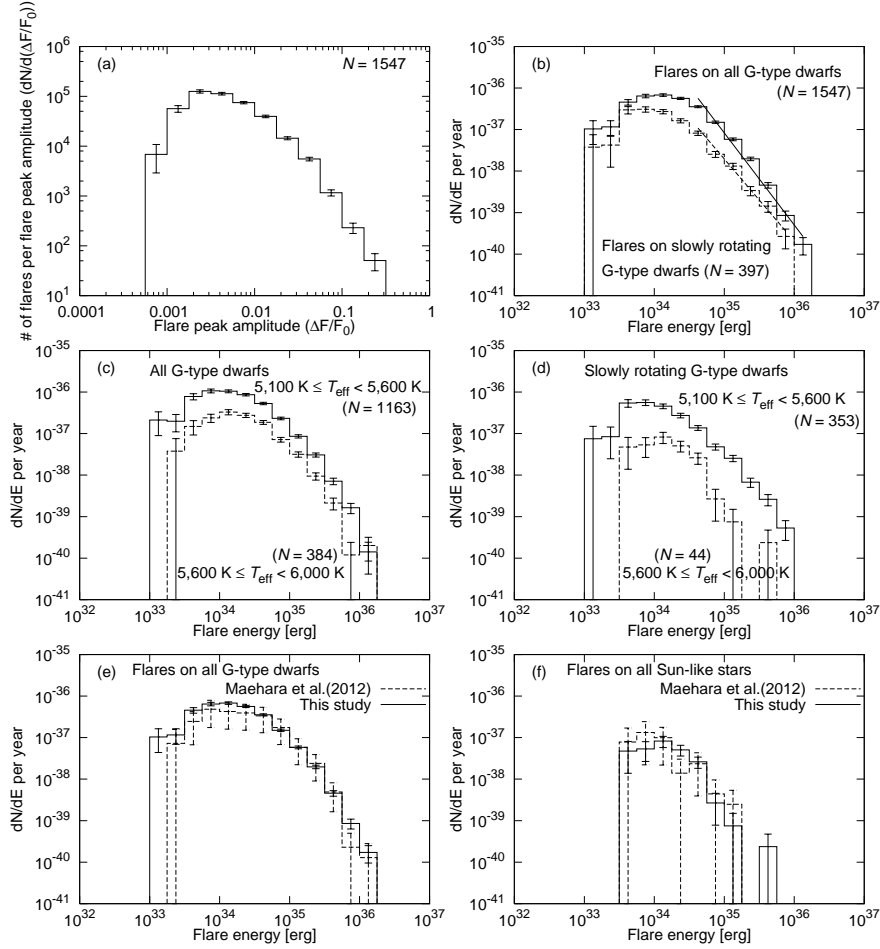}
\caption{Frequency distribution of superflares on G-type dwarfs 
(5100 K $\leq \teff <$ 6000 K, and $\logg > 4.0$).
(a) Number distribution of superflares as a function of the observed amplitude.
The number of flares ($N$) is 1547. 
Note that the y-axis values have been divided by the bin width.
(b) Frequency distribution of flares as a function of the flare energy.
Solid and dashed lines show distributions on all stars and 
slowly rotating stars (the timescale of the brightness variation $> 10 d$).
The power-law index of all G-type dwarfs and slowly rotating G-type dwarfs
is $\sim2.2$ and $\sim2.0$ respectively.
(c) Same as (b). but for flares on all lower-temperature stars
(5,100 K $\leq \teff <$ 5,600 K ; solid line) and
all high temperature stars (5,600 K $\leq \teff <$ 6,000 K ; dashed line).
(d) Same as (c) but only for flares on slowly rotating stars.
(e), (f) Companions of the results in this study and \citet{maehara2012}.
All errors are estimated to be a square root of the number of flares in each bin.
}
\label{fig:panels}
\end{figure}

Figure \ref{fig:panels} represent the occurrence frequency distributions of superflares.
The configuration of this figure is basically the same as Figure 2 of \citet{maehara2012} 
but the number of flares is more than 4 times that of flares analyzed by \citet{maehara2012}. 
In  Figure \ref{fig:panels}(a) we show the number distribution of flares as a function of observed flare amplitude. 
The number of flares (N) is 1547, and the errors are estimated from the square root of the event number in each bin. 
Error bars from small number statistics are given by binomial distribution \citep{gehrels1986},
the error bars in figures, therefore, are overestimated when the number of flares is not enough large.

The average occurrence frequency of superflares can be estimated from the number of observed superflares, 
the number of observed stars and the length of the observational period. 
For example, in the case of Sun-like stars
44 superflares were detected from the data of about 14000 stars over 500 days. 
Hence, the occurrence frequency of superflares is $2.3 \times 10^{-3}$ flares per year per star, 
which corresponds to a superflare occurring on a star once every $440_{-60}^{+80}$ yr.
We show the frequency distributions of flares 
as a function of flare energy in Figure \ref{fig:panels}(b), (c) and (d), 
but the sets of superflare generating stars for each curve are different.
The vertical axis indicates the number of superflares per star per year per unit energy. 
We show the distribution of all superflares (solid line) and superflares on slowly rotating (longer than 10 days period) stars (dashed line) in the panel (b),
superflares on all G-type dwarfs whose effective temperature ($\teff$) is 5,100 K $\leq \teff <$ 5,600 K (solid line) 
and $5,600 K \leq \teff < 6,000 K$ (dashed line) in the panel (c), 
and slowly rotating G-type dwarfs of the same temperature range as (c) in the panels (d).
We show comparisons of this study and \citet{maehara2012} in the panels (e) and (f).
Results of this study basically agree with the results in \citet{maehara2012}, 
and we can see the errors become smaller.
It is seen from Figure \ref{fig:panels}(b) that the occurrence frequency of superflares
in all G-type dwarfs show a power-law distribution 
$$dN/dE \propto E^{-\alpha}$$
with $\alpha \simeq 2.2$ for all G-type dwarfs and $\alpha \simeq 2.0$
for flares on slowly rotating G-type dwarfs. It is interesting to note
that these distributions are quite similar to those for solar flares \citep[e.g.,][]{aschwanden2000} and 
stellar flares on red dwarfs \citep[e.g.,][]{shakhovskaia1989}.

We further found from Figures \ref{fig:panels}(b)-(d) that 
(1) the frequency of superflares on slowly rotating stars 
is smaller than that for all G-type dwarfs, and also 
(2) the frequency of superflares on hot G-type dwarfs ($5,600 < \teff < 6,000$ K)
is smaller than that for cool G-type dwarfs ($5,100 < \teff < 5,500$ K).
These are basically the same as those found in \citet{maehara2012}.
The number of superflares on Sun-like stars is 36 on  13 stars, which
is much more than 14 superflares on 10 Sun-like stars in \citet{maehara2012}.
The occurrence frequency distribution of superflares on these Sun-like stars
is again similar to that for our Sun, and the occurrence frequency of superflares
with energy of $10^{34}$ erg is once in 800 years and that of $10^{35}$ erg
is once in 5000 years. 

\subsection{Hot Jupiter}

According to Kepler candidate planet data explorer \citep{batalha2012}, 
2321 planets have been found in 1790 stars among 156453 stars.
Hence the probability of finding exoplanets on stars is about 1 percent.
\citet{howard2012} showed that the probability of finding hot Jupiter
was 0.5 percent. 
However all our superflare stars (279 G-type dwarfs) do not have 
hot Jupiter according to above data explorer. 

For a G-type dwarf with a hot Jupiter, the probability of a transit of the
planet across the star is about 10 percent averaged over all possible
orbital inclinations \citep{kane2008}. 
If all our 279 superflare stars are caused by hot Jupiter as suggested
by \citet{rubenstein2000}, Kepler should detect 28 of them from transits. 
However, we did not find any hot Jupiter around our superflare stars
as mentioned above.
This suggests that hot Jupiters associated with superflares are rare.

\subsection{Occurrence Frequency of Superflares on the Most Active Stars}

\begin{figure}
\epsscale{1}
\plotone{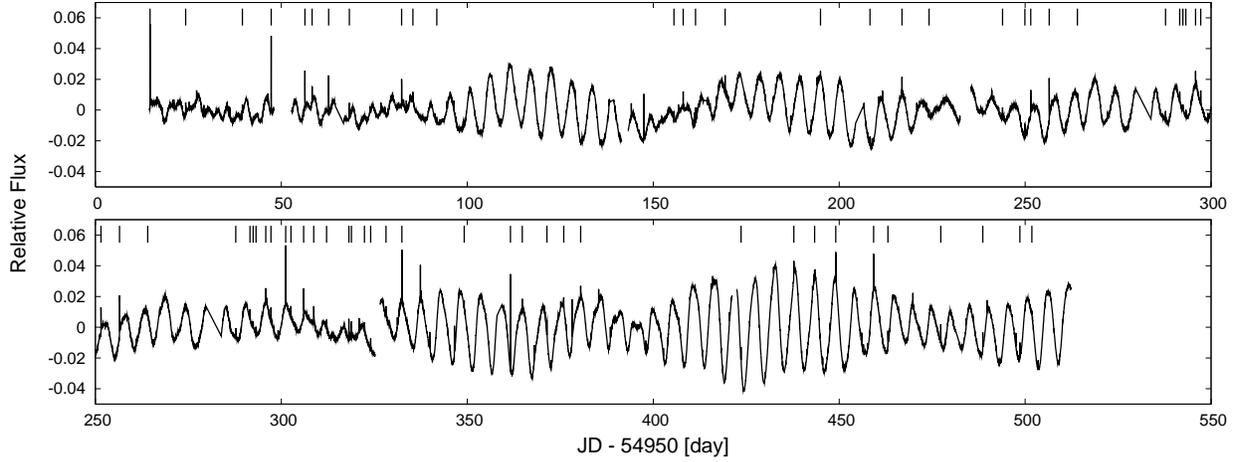}
\caption{Long term stellar brightness variations of KIC10422252.
Vertical short lines indicate the peak time of the superflares we detected.
The vertical axis show the brightness variations relative to the average brightness.}
\label{fig:longst}
\end{figure}

Figure \ref{fig:longst} represents the lightcurve of the most active superflare star (KIC10422252). 
This star exhibits 57 superflares in about 500 days, 
and hence the superflare occurrence frequency is more than once in 10 days.
The variation of the stellar brightness contains modulation of the frequency which is different from but near the main frequency. 
Light curves of that sort is typically shown by the starspots system 
with differential rotation \citep{frasca2011}.

\begin{figure}
\epsscale{1}
\plotone{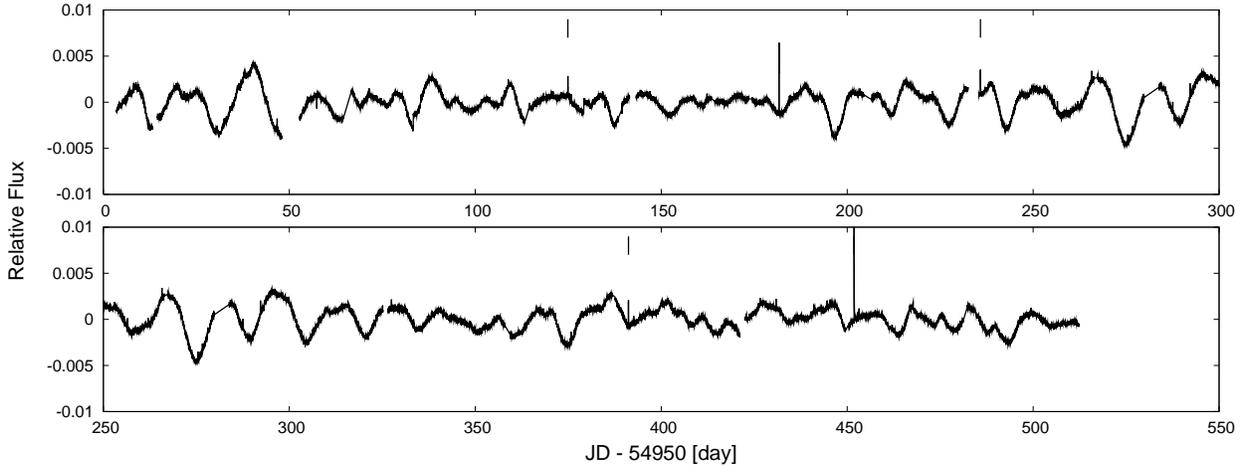}
\caption{Same as Figure \ref{fig:longst}, but for KIC10471412.
This star is similar to the Sun from a viewpoint of surface temperature,
surface gravity, and rotational period. 
The rotational period of this star is estimated to be 17.1 days from 
the brightness variation of the light curve.
}
\label{fig:longsl}
\end{figure}

Figure \ref{fig:longsl} is the same as Figure \ref{fig:longst} but for the KIC10471412.
This star is an active Sun-like star, which produced 4 superflares 
in the whole observation period (500 days).
This occurrence frequency of superflare corresponds to once in 100 days.

\begin{figure}
\epsscale{1}
\plotone{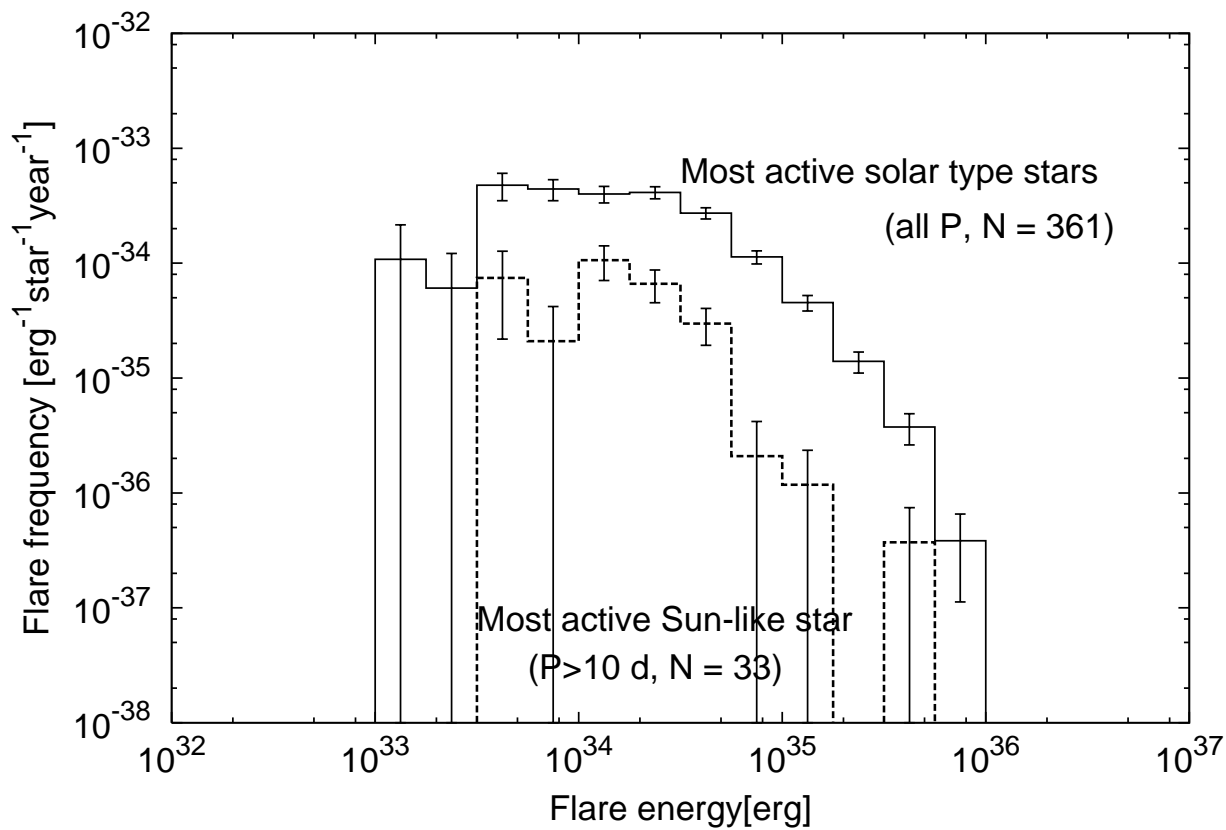}
\caption{Same as Figure \ref{fig:panels}, but for most active G-type dwarfs (solid line), and most active Sun-like stars (dashed line).
The most active G-type dwarf is defined as a G-type dwarf 
having a superflare occurrence frequency larger than once in $\sim$10 d. 
The most active Sun-like star is defined as a Sun-like star
which exhibited more than 1 superflare in the whole observation period.}
\label{fig:ma}
\end{figure}

Figure \ref{fig:ma} shows the flare occurrence frequency distribution of 
the most active G-type dwarfs and Sun-like stars.
The solid line correspond to average frequency of superflares on
9 most active G-type dwarfs, 
whose superflare occurrence frequency is more than once in 10 days.
The dashed line show the distribution for Sun-like stars that
showed more than one superflare in the whole observing period (about 500 days).
Error bars are estimated from the square root of the numbers of superflares in each bin.

It is interesting to note that the frequency of superflares with $\sim 10^{34}$ erg in the dM4.5e star YZ CMi
is once in a month \citep{lacy1976, kowalski2010}, which is comparable to those of the most active G-dwarfs and Sun-like stars (once in 10 - 100 days).

\section{Discussion}

\subsection{Occurrence Frequency Distribution of Superflares}

\begin{figure}
\epsscale{1}
\plotone{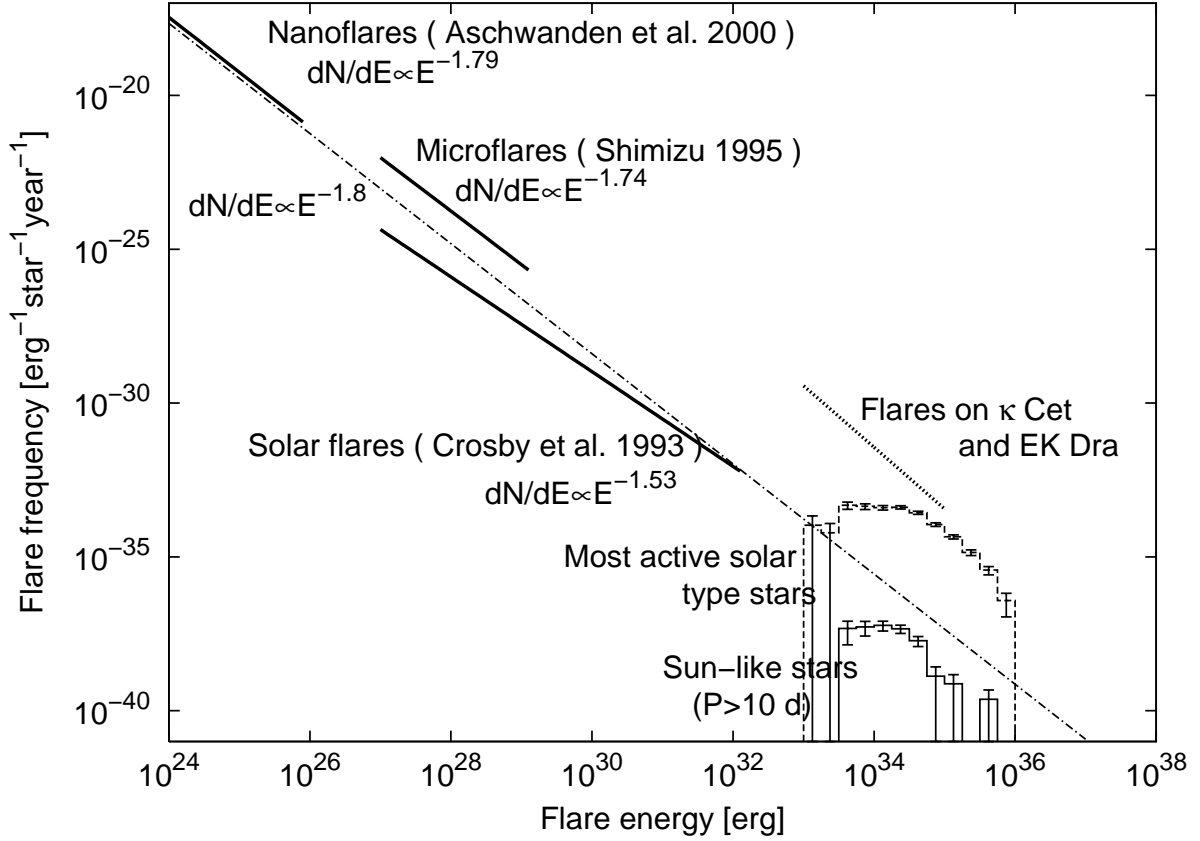}
\caption{Occurrence frequency of superflares on G-type dwarfs and of solar flares. 
The solid-line histogram shows the frequency distribution 
of superflares on Sun-like stars, while the occurrence frequency of 
superflares on most active G-type dwarfs (with flares more than once in about 10 days)
is illustrated by the dashed line histogram. 
The error bars in the histogram are estimated from the square root of event number in each bin. 
The solid lines indicate the power-law distribution of solar flares 
observed in EUV \citep{aschwanden2000}, soft X-rays \citep{shimizu1995}, 
and hard X-rays \citep{crosby1993} respectively. 
It is interesting that superflares in Sun-like stars, solar flares, microflares, 
and nanoflares are roughly on the same power-law line with an index of -1.8 (thin dash-dotted line) 
for a wide energy range from $10^{24}$ erg to $10^{35}$ erg.}
\label{fig:dist}
\end{figure}

We shall compare the occurrence frequency distribution of 
superflares with those of solar flares 
and stellar flares on active G-type dwarfs discussed by \citet{schrijver2012}
in Figure \ref{fig:dist}.
The solid-line histogram shows the flare occurrence frequency distribution of superflares on 
G-type dwarfs and dashed histogram correspond to the frequency distribution of superflares on 
most active G-type dwarfs. 
The error bars in the histogram represent the square root of event number in each bin.
Solid lines indicate the frequency distribution of nanoflares observed in 
EUV \citep{aschwanden2000}, microflares in 
soft X-rays \citep{shimizu1995}, and solar flares in hard X-rays \citep{crosby1993} respectively.
Thin dot-dashed line shows a power-law line with 
the distribution $dN/dE \propto E^{-\alpha}$ with index $\alpha \sim 1.8$.
The thick dotted line corresponds to the frequency distribution of stellar flare on
active G-type dwarfs shown in \citet{schrijver2012} ($\kappa$ Cet, EK Dra, and 47 Cas).
The frequency of flares on these G-type dwarfs is comparable to that of   
most active G-type dwarfs in our data. 
The maximum energy of these stellar flares is lower than that of superflares 
on the most active G-type dwarfs.

It is quite interesting to see that solar flares and superflares on Sun-like stars are
roughly on the same power-law line. This suggests that the mechanism of superflares is 
similar to that of solar flares and microflares, because white-light flare energy 
is roughly in proportion to EUV and X-ray flare energy 
(but see Fig \ref{fig:isii2} for the case $F_{\rm white-light} \propto F_{\rm X-ray}^{0.65}$).

In Figure \ref{fig:isii}, we compare the frequency distribution of superflares on  the most active
Sun-like stars (the same curve as the dashed line of Figure \ref{fig:ma})  with those of solar flares and average Sun-like stars.
It is found that the frequency of superflare occurrence in the most active 
Sun-like stars is about 1000 times larger than that of average Sun-like stars.
In Figure \ref{fig:isii}, we also added the flare frequency in solar maximum (in 2001) 
and minimum (in 2008) using the statistics of solar flares based on
GOES X-ray flux classification
(X-class, M-class, C-class flares; see Table \ref{tab:goes}, \citet{ishii2012}). 
Here we assumed that the bolometric flux of flare (${\rm F}_{bol}$) (70\% white-light; \citet{kretzschmar2011}) 
is in proportion to the GOES X-ray flux (${\rm F}_{GOES}$). 
\footnote{See Figure \ref{fig:method}2 and Appendix B for the case
of flare energy estimate using \citet{kretzschmar2011}'s empirical
relation between the total solar irradiance and GOES X-ray flux of a solar flare.}
It is seen that the flare frequency in the solar maximum 
is larger than that in the solar minimum by about a factor of 200.
It is also found that 
the superflare frequency curve in the most active Sun-like stars
is well above that of solar flares during the maximum solar cycle are
We can now understand from this figure that the difference in54the activity level of the Sun (or sunspot number or
total magnetic flux) leads to the difference of solar flare frequency in 
the different phase of the solar activity cycle. 

This suggests that the total magnetic flux
of the most active Sun-like stars is larger than that of the solar maximum.
Since the total magnetic flux during the solar maximum and minimum 
are about $10^{24}$ Mx and  $10^{23}$ Mx
 \citep[e.g.,][]{solanki2002, vieira2010},  the total magnetic
flux in the most active Sun-like stars may be estimated to be
$3 \times 10^{24} \sim 10^{25}$ Mx.  

Further, we show the frequency of stellar flares on 
$\kappa$ Cet with a thick dotted line in Figure \ref{fig:isii}.
$\kappa$ Cet is a G5 type main sequence star 
($\teff = 5520 {\rm K}, v\sin i = 3.9 {\rm km/s}, P_{rot}=9.4 {\rm d}$) \citep{schrijver2012, audard2000},
and hence is slightly cooler than Sun-like stars.
The frequency of flares on $\kappa$ Cet is larger than that of 
most active Sun-like stars by a factor of 100, although
the maximum energy of flares on $\kappa$ Cet ($10^{34}$ erg) is lower than that of flares 
on most active Sun-like stars ($5 \times 10^{35}$ erg).

Flare frequency of Sun-like stars is assumed to be determined by the average magnetic activity level of these stars.
In Figure \ref{fig:isii}, it is interesting to see that flare occurrence frequency distribution of 
average of all Sun-like stars is located between the solar maximum and minimum.
This suggests that average activity level of all Sun-like stars  
are comparable to that of the Sun. 
It is suggested from Figure \ref{fig:isii} 
that Sun-like stars have hyper active cycle, when the star exhibits 1000 times more 
flares than the average of Sun-like stars.
The number of all Sun-like stars observed by the Kepler is 14410 and
the number of the hyper active stars we detected was 6.
The fraction of stars showing hyper active cycle is, therefore, of the order of $10^{-4}$.

\subsection{Dependence of Flare Frequency on Rotation of Superflare Stars}

Rapidly rotating stars (rotational period : $P < 10 d$) tend to exhibit superflares 
more frequently than slowly rotating stars ($P > 10 d$) (see Table \ref{tab:superflare}, \ref{tab:allstars}).
As mentioned in the Section 2.2, the flare selection criteria, and therefore the detection completeness of flares, 
depends on the rotation period of the star.
Our result shows, however, that rapidly rotating stars show higher flare occurrence frequency than slowly rotating stars.
This trend cannot be affected by the nonuniformity of the detection completeness.  
The rotation period is correlated with the chromospheric activity, which is known to be an indicator of the magnetic activity of the stars \citep{noyes1984}, 
and the rapidly rotating stars have higher magnetic activity than slowly rotating stars \citep{pallavicini1981}. 
According to the dynamo theory of magnetic field generation, magnetic activity results from the interaction between 
rotation and convection \citep{parker1979}, and the rapid rotation can cause the high magnetic activity. 
Our result implies that rapidly rotating stars with higher magnetic activity can cause more frequent superflares.

The rotation period of a star is also known to be related to the stellar age, 
and younger stars show more rapid rotation \citep{skumanich1972,barnes2003}. 
Our findings suggest that superflares occur more frequently on the young G-type dwarfs (i.e. rapidly rotating stars). 
Moreover, on G-type dwarfs similar in age to the Sun, superflares occur less frequently, though the maximum energy of superflares is nearly independent of
rotation period \citep{maehara2012}.  

The relation between properties of superflares and rotation
period of superflare stars will be analyzed in detail in our 
paper \citep{notsu2013b}.

\subsection{Superflares on the Sun ?}

It has been pointed out that there is no record of solar superflares over the past 2,000 yr \citep{schaefer2000}. 
According to the measurement of the impulsive nitrate events in polar ice, 
the largest proton flare event during the past 450 yr is the Carrington event \citep{shea2006}, which occurred on 1 September 1859 \citep{carrington1859}. 
The total energy released in this flare was estimated to be of order $10^{32}$ erg \citep{tsurutani2003}, 
which is only 1/1,000 of the maximum energy of flares on slowly rotating Sun-like stars we detected.
Our criterion of rotation period for Sun-like stars is more than 10 days, 
and the average of rotation period of Sun-like superflare stars is 12.7 d.
This rotation period corresponds to rotation velocity of $\sim4$ km/s.
According to the rotation-age relation for single stars in \citet{ayres1997},
this rotation velocity corresponds to an age of $\sim1$ Gyr.
The rotational velocity of the Sun is $\sim2$ km/s
(rotational period is 25 days) 
and the age of our Sun is 4.6 Gyr.
Do superflares really occurr on Sun-like stars more similar to the Sun
which have a rotational period longer than 25 days?
Table \ref{tab:sunlike} shows 4 superflares 
($5.0\times10^{34},~2.4\times10^{34},~9.9\times10^{33}$ 
and $4.4\times10^{34}$ erg) occurred on Sun-like stars with 
surface temperature $5600~{\rm K}<T_{\rm eff}<6000$ K and rotational 
period longer than 25 days.
The number of such stars is about 5000 in our sample. 
This implies that superflare whose total energy is of the order of 
$10^{34}$ erg might occur on our Sun once in 
$\sim2000$ years, although more detailed analysis is needed.

It has also been proposed that hot Jupiters have an important effect on stellar magnetic activity \citep{rubenstein2000,cuntz2000,ip2004}
and that superflares occur only on G-type dwarfs with hot Jupiters. 
However, there is no hot Jupiter in the Solar System. 
For these reasons, it was suggested that a superflare on the Sun was extremely unlikely \citep{schaefer2000,rubenstein2000}.

We found, however, no hot Jupiter around the 279 G-type dwarfs with superflares (see sec. 3.3).
This suggests that hot Jupiter is not a necessary condition for superflares.
Recent theoretical research \citep{shibata2013} also suggested that G-type dwarfs can store magnetic energy, 
which is enough for causing superflare, in overshoot layer.
Moreover, \citet{miyake2012} reported an occurrence of a 
energetic cosmic ray
event in 8th century recorded in a tree ring of Japanese cedar trees.
There is a possibility that this event was produced by a superflare 
(with energy of $\sim 10^{35}$ erg) on our Sun.

In this paper, we found 1547 superflares on 279 G-type dwarfs 
in 500 days Kepler data. Using these superflares, we calculated 
superflare occurrence frequency distributions as a function of 
total superflare energy. This shows power-law distributions and the
power-law index was comparable with that of solar flares ($\sim 2$), 
and the average occurrence frequency of superflares with 
energy of $10^{34}-10^{35}$ erg is once in 800-5000 years.
According to Kepler candidate planet data explorer, there was no
host star of hot Jupiter in all superflare stars.
We showed a comparison of flare occurrence distributions of solar 
maximum, minimum, and superflare stars, and pointed out the existence
of hyper active cycle of G-type dwarfs.
We found hyper active G-type superflare stars, which exhibit superflares more
than once in 10 days.
We found 36 superflares on Sun-like stars, and this is much more
than 14 superflares on Sun-like stars discovered by \citet{maehara2012}.
Nevertheless, this number is still not
enough to derive accurate statistical properties of superflares
on Sun-like stars. This is important for discussion of the
possibility of superflares on our present Sun. 
Hence it will be important to continue to observe Sun-like stars
for a longer time to increase the number of superflares on 
Sun-like stars.  It will be also important to get short time
(1 min) cadence data to obtain information on small superflares
(with energy of $10^{33} - 10^{34}$ erg) which bridge the gap
between superflares on Sun-like stars and largest solar flares
$3 \times 10^{32}$ erg. 
The Kepler only obtain photometric data. 
Rotational periods and areas of star spots calculated from them 
are indirect estimate based on some presumptions.
In order to get direct evidence of slow rotation, large spot, high activity,
and binarity,
high dispersion spectroscopic observation is necessary.
We have observed some superflare stars with the HDS on Subaru telescope.
The results from these observations will be reported elsewhere \citep{notsu2013a}.

\acknowledgments
We are grateful to Prof. Kazuhiro Sekiguchi (NAOJ) for useful suggestions. 
We also thank the anonymous referee for helpful comments. 
Kepler was selected as the tenth Discovery mission. Funding for this mission 
is provided by the NASA Science Mission Directorate. The data presented 
in this paper were obtained from the Multimission Archive at STScI. 
This work was supported by the Grant-in-Aid from the Ministry of 
Education, Culture, Sports, Science and Technology of Japan (No. 25287039).

\appendix
\section{Detailed Light Curve}
We used light curves of long time corrected flux (time resolution is about 30 min) to detect superflares.
We can obtain the light curve of flare with short time cadence data (1min) in some objects.
We compare the estimated energy from long time data and short time data. (Details are shown in Sec.2)
Figure \ref{fig:ap1} shows the comparisons of the light curve of flare with long data and short data in KIC11610797.
The estimated energy of flare from long data is $1.4 \times 10^{35}$ erg and from short data is $1.4 \times 10^{35}$ erg, respectively.
Figure \ref{fig:ap2} is the same as Figure \ref{fig:ap1}, but for KIC965268.
The energy of flare estimated from long data is $1.5 \times 10^{34}$ erg and from short data is $1.9 \times 10^{34}$ erg.
Those values are in agreement within the error.
Hence, we conclude the flare energies estimated from long time cadence data are reliable.

\section{Estimating total flare energy from GOES X-ray flux
based on Kretzschmar's relation}

\begin{figure}
\epsscale{1}
\plotone{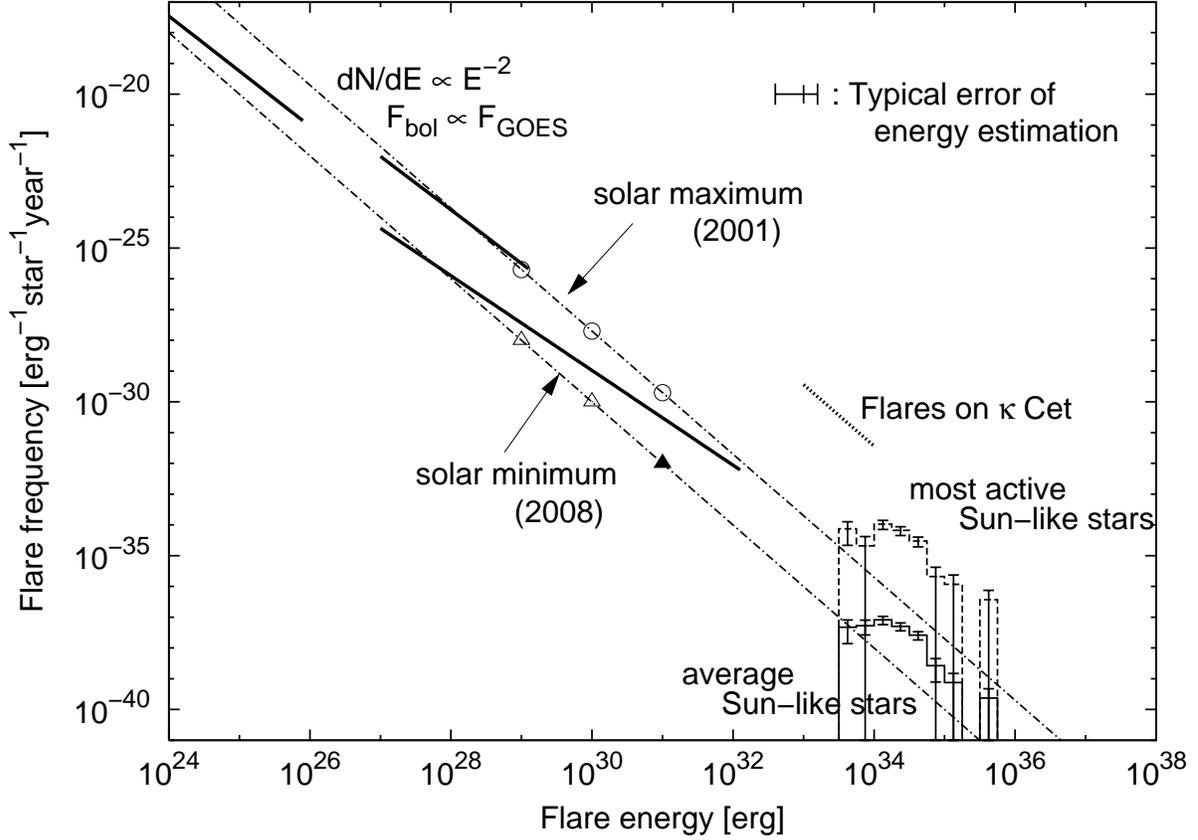}
\caption{Flare occurrence frequency as a function of the flare energy
in light on the stellar activity cycle.
Solid lines correspond to power-law distributions of solar flares
, which are the same lines as in Figure \ref{fig:dist}.
Distribution histograms represent the flare occurrence frequency of superflares on all Sun-like stars (solid curve), and
on Sun-like stars with more than one superflare (dashed curve).
The bar at the upper right of this figure shows a typical error of our energy estimate.
Dash-dotted lines correspond to power-law
distributions estimated from numbers of solar flares observed by GOES.
The flux in bolometric flux of flare ($F_{\rm bol}$) is estimated to be in proportion to the GOES X-ray Flux ($F_{\rm GOES}$).
We show the case of $F_{\rm bol} \propto F_{\rm GOES}^{0.65}$ in Figure \ref{fig:isii2}. 
Open circles show the occurrence frequency of C, M, X class solar flares during the solar maximum, while
frequency during the solar minimum is shown by triangles. 
No X class solar flare was observed in the solar minimum (2008). 
We estimated the occurrence frequency of X class flare in the solar minimum,
and the estimated frequency is shown by the closed triangle.}
\label{fig:isii}
\end{figure}

\begin{figure}
\epsscale{1}
\plotone{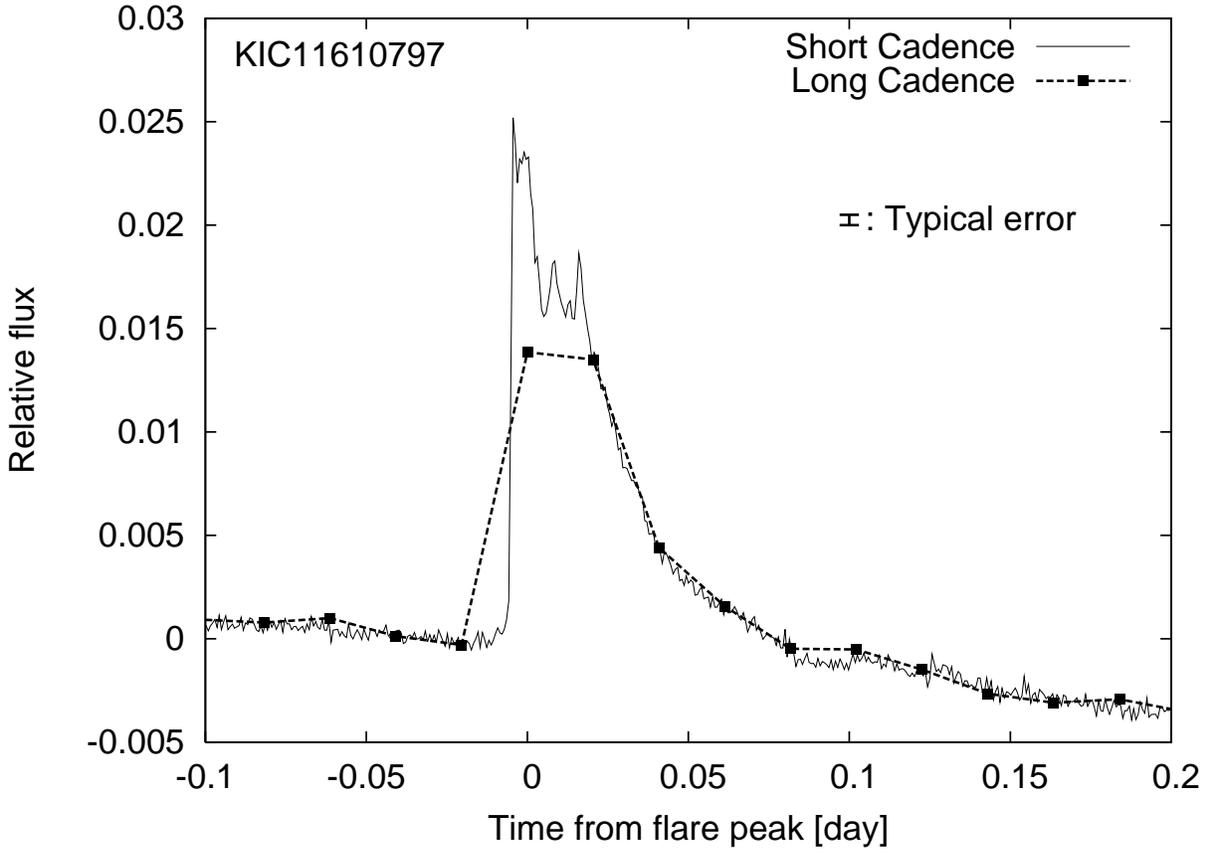}
\caption{Light curves of a superflare on KIC11610797. Dashed solid lines show the light curves of long and short time cadence data respectively.}
\label{fig:ap1}
\end{figure}

\begin{figure}
\epsscale{1}
\plotone{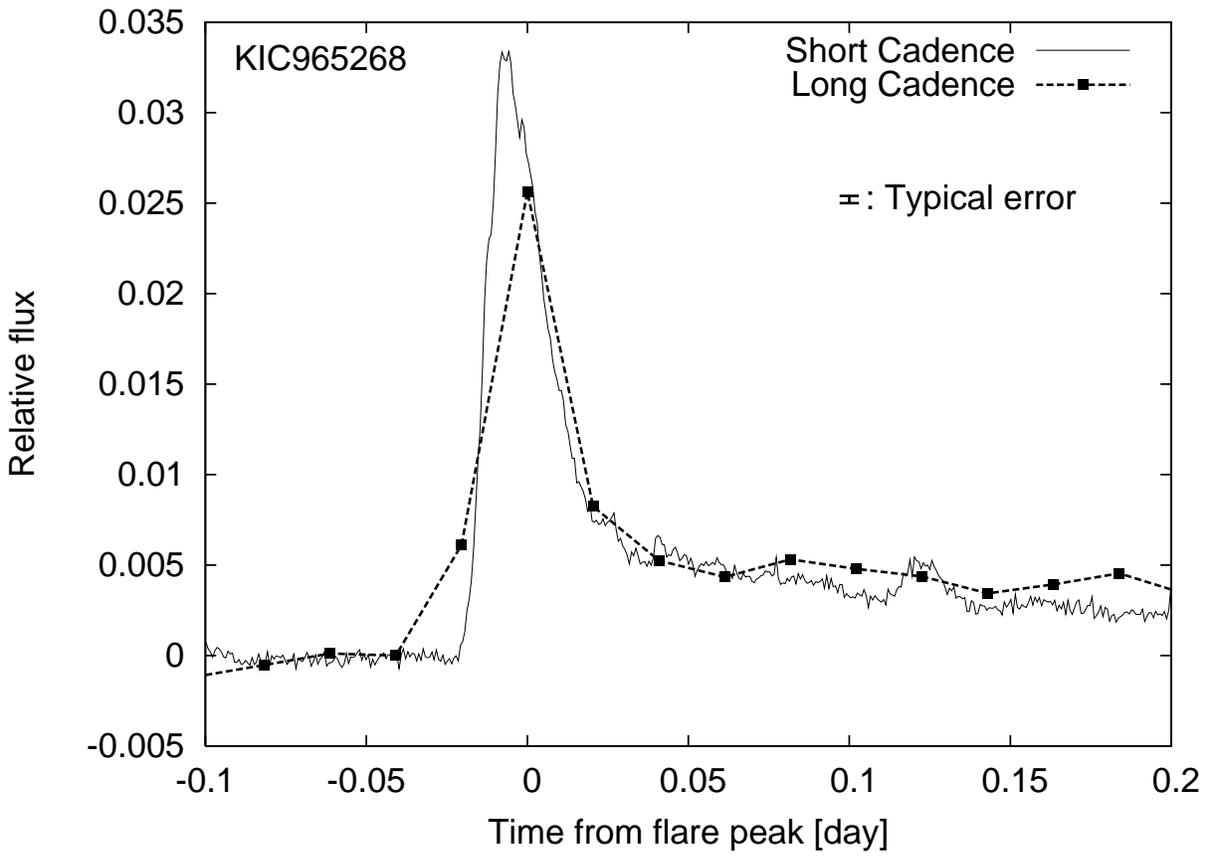}
\caption{Same as Figure \ref{fig:ap1}, but for KIC9652680.}
\label{fig:ap2}
\end{figure}

\begin{figure}
\plotone{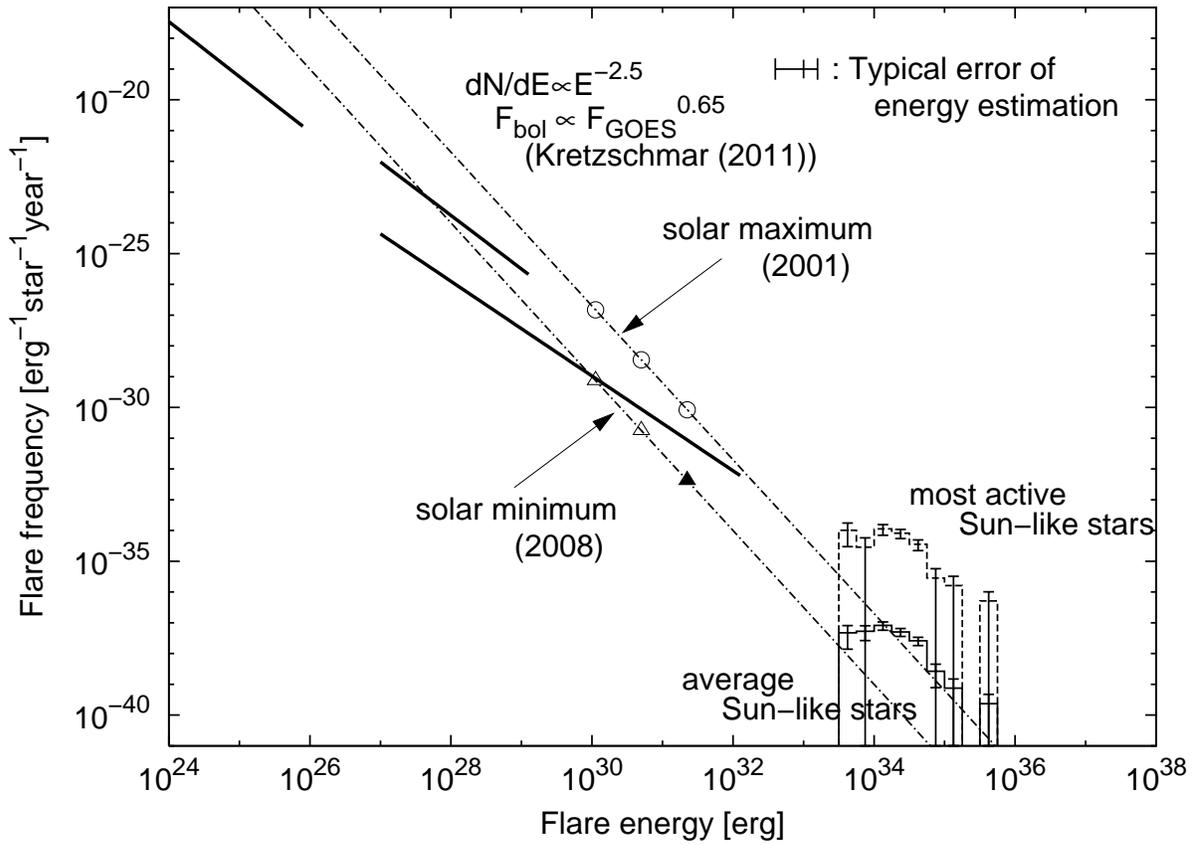}
\caption{Same as Figure \ref{fig:isii}, but the total solar irradiance is estimated to be in proportion to the GOES X-ray Flux to the power of 0.65 \citep{kretzschmar2011}.}
\label{fig:isii2}
\end{figure}

\citet{kretzschmar2011} derived the empirical relation 
between the bolometric flux of flare and GOES X-ray flux as
$$ F_{bol} = 2.4 \times 10^{12} F_{GOES}^{0.65} . $$
If we use this relation to derive the total flare energy
from GOES X-ray flux assuming the total energy
of X10 class flare  is $10^{32}$ erg, then the 
total energy of X-class, M-class, C-class flares
become $10^{31.4}$ erg, $10^{30.7}$ erg, $10^{30.1}$ erg,
respectively.  Figure \ref{fig:isii2} is the same as Figure \ref{fig:isii} but for
the case of the flare energy estimate using Kretzschmar's relation.
In this case, we clearly see disagreement between the observed
frequency of nanoflares and those based on
Kretzschmar's relation.

\begin{table}[htpb]
\begin{center}
\caption{Length of the observation period during each quarter and the number of
G-type dwarfs.}
\begin{tabular}{ccccc}
\hline
Quarter&N\tablenotemark{\dag}&$\tau$[days]\tablenotemark{\ddag}&Start Date[UT]&End Date[UT]\\
\hline
0 & 9511 & 10 & 02-May-2009 & 11-May-2009\\
1 & 75598 & 33 & 13-May-2009 & 15-Jun-2009\\
2 & 82811 & 89 & 20-Jun-2009 & 16-Sep-2009\\
3 & 82586 & 89 & 18-Sep-2009 & 16-Dec-2009\\
4 & 89188 & 90 & 19-Dec-2009 & 19-Mar-2010\\
5 & 86248 & 95 & 20-Mar-2010 & 23-Jun-2010\\
6 & 82052 & 90 & 24-Jun-2010 & 22-Sep-2010\\
\hline
\label{data_summary}
\end{tabular}
\tablenotetext{\dag}{Number of G-type dwarfs.} 
\tablenotetext{\ddag}{~Length of the observation period during each quarter.} 
\label{tab:nstar}
\end{center}
\end{table}

\begin{table}[htpb]
\begin{center}
\caption{Number of flares, flarestars, G dwarfs}
\begin{tabular}{ccccc}
\hline
$\log {\rm P}$ & $N_{\rm flare}$ & $N_{\rm flarestar}$ & $N_{\rm star}$ & $N_{\rm star,old}$ \tablenotemark{\dag}\\
\hline
-1.0 & 2 & 1 & 38 & 34 \\
-0.8 & 0 & 0 & 331 & 260 \\
-0.6 & 1 & 1 & 219 & 156 \\
-0.4 & 9 & 5 & 251 & 159 \\
-0.2 & 35 & 12 & 334 & 215 \\
0.0 & 34 & 12 & 393 & 299 \\
0.2 & 65 & 20 & 422 & 428 \\
0.4 & 127 & 29 & 656 & 870 \\
0.6 & 120 & 27 & 1364 & 2000 \\
0.8 & 75 & 16 & 3562 & 5549 \\
1.0 & 28 & 8 & 9003 & 10248 \\
1.2 & 102 & 13 & 19039 & 9338 \\
1.4 & 27 & 3 & 30131 & 4919 \\
1.6 & 0 & 0 & 17063 & 326 \\
\hline
\end{tabular}
\tablenotetext{\dag}{Period is calculated from the old data.} 
\label{tab:period}
\end{center}
\end{table}

\begin{table}
  \begin{center}
    \caption{Parameters of stars illustrated in the figures}
    \begin{tabular}{cccccc}
      \tableline
      KeplerID & $\teff$ \tablenotemark{a}(K)& $\logg$ \tablenotemark{b} & $R/R_{\odot}$ \tablenotemark{c}& KPmag \tablenotemark{d} &$P_{\rm rot}(day)$ \tablenotemark{e}  \\
      \tableline
      4245449 & 5761 & 4.0 & 1.8 & 12.6 & 2.1 \\
      4543412 & 5907 & 4.3 & 1.3 & 11.2 & 2.2 \\
      6865484 & 5688 & 4.4 & 1.1 & 13.8 & 11.2 \\
      9574994 & 5925 & 4.4 & 1.1 & 15.1 & 12.3 \\
      10120296 & 5490 & 4.4 & 1.1 & 12.9 & 3.9 \\
      10422252 & 5118 & 4.2 & 1.3 & 13.6 & 5.2 \\
      10471412 & 5771 & 4.1 & 1.6 & 13.5 & 15.1 \\
      10524994 & 5747 & 4.5 & 1.0 & 15.3 & 12.0 \\
      12354328 & 5115 & 4.4 & 1.0 & 14.7 & 0.8 \\
      \tableline
    \end{tabular}
    \tablenotetext{a}{Effective temperature}
    \tablenotetext{b}{Surface gravity}
    \tablenotetext{c}{Stellar radius in unit of the solar radius}
    \tablenotetext{d}{Kepler magnitude \citep{brown2011}}
    \tablenotetext{e}{Stellar rotation period estimated from brightness variation}
    \label{tab:starparameter}
  \end{center}
\end{table}

\begin{table}
  \begin{center}
    \caption{Annual Variation of 
Number of Solar Flares observed by GOES}
    \begin{tabular}{cccc}
      \tableline
      Year & $N_X$ \tablenotemark{\dag}& $N_M$ \tablenotemark{\dag}& $N_C$ \tablenotemark{\dag}\\
      \tableline
      1989 & 59 & 620 & 1929 \\
      1990 & 16 & 273 & 2262 \\
      1991 & 54 & 590 & 2653 \\
      1992 & 10 & 202 & 1922 \\
      1993 & 0 & 74 & 1142 \\
      1994 & 0 & 25 & 336 \\
      1995 & 0 & 11 & 148 \\
      1996 & 1 & 4 & 81 \\
      1997 & 3 & 21 & 286 \\
      1998 & 14 & 94 & 1188 \\
      1999 & 4 & 170 & 1854 \\
      2000 & 17 & 215 & 2223 \\
      2001 & 21 & 310 & 2101 \\
      2002 & 12 & 219 & 2323  \\
      2003 & 20 & 160 & 1316  \\
      2004 & 12 & 122 & 912  \\
      2005 & 18 & 103 & 599  \\
      2006 & 4 & 14 & 174  \\
      2007 & 0 & 10 & 73  \\
      2008 & 0 & 1 & 8 \\
      2009 & 0 & 0 & 28 \\
      2010 & 0 & 23 & 170 \\
      2011 & 8 & 111 & 1200 \\
      \tableline
    \end{tabular}
    \tablenotetext{\dag}{Number of X, M and C class flares}
    \label{tab:goes}
  \end{center}
\end{table}

\clearpage

\begin{table}
\begin{center}
\caption{Number of Superflares and Superflare Stars}
\begin{tabular}{ccccccccc}
\hline
          & \multicolumn{2}{c}{Slow\tablenotemark{\dag}} & & \multicolumn{2}{c}{Fast\tablenotemark{\dag}} & & \multicolumn{2}{c}{Total\tablenotemark{\dag}}\\
\hline
$\teff$ & $N_{\rm flare}$\tablenotemark{\ddag} &  $N_{\rm fstar}$\tablenotemark{\ddag} & & $N_{\rm flare}$\tablenotemark{\ddag} & $N_{\rm fstar}$\tablenotemark{\ddag} & & $N_{\rm flare}$\tablenotemark{\ddag} & $N_{\rm fstar}$\tablenotemark{\ddag} \\
\hline
5100-5600 & 353 & 50 & & 810 & 133 & & 1163 & 183 \\
5600-6000 & 44 & 19 & & 340 & 77 & & 384 & 96 \\
\hline
          & 497 & 69 & & 1150 & 210 & & 1547 & 279 \\
\hline
\label{tab:superflare}
\end{tabular}
\tablenotetext{\dag}{Categorization by the $P$ (stellar rotation period estimated from lightcurve). $P>10d$ (Slow), $P<10d$ (Fast) and the total of them (Total).} 
\tablenotetext{\ddag}{Number of superflare ($N_{\rm flare}$) and number of flare stars in the category ($N_{\rm all}$).}
\end{center}
\end{table}

\begin{table}
\begin{center}
\caption{Number of Super Flare Stars and Observed Stars}
\begin{tabular}{cccccccccccc}
\hline
          & \multicolumn{3}{c}{Slow\tablenotemark{\dag}} & & \multicolumn{3}{c}{Fast\tablenotemark{\dag}} & & \multicolumn{3}{c}{Total\tablenotemark{\dag}}\\
\hline
$\teff$ & $N_{\rm fstar}$\tablenotemark{\ddag} &  $N_{\rm all}$\tablenotemark{\ddag} & $f_{\rm fstar}$\tablenotemark{\ddag} & & $N_{\rm fstar}$\tablenotemark{\ddag} & $N_{\rm all}$\tablenotemark{\ddag} & $f_{\rm fstar}$\tablenotemark{\ddag} & & $N_{\rm fstar}$\tablenotemark{\ddag} & $N_{\rm all}$\tablenotemark{\ddag} & $f_{\rm fstar}$\tablenotemark{\ddag}\\
\hline
5100-5600 & 50 & 14026 & 0.0036 & & 133 & 1281 & 0.104 & & 183 & 15307 & 0.012\\
5600-6000 & 19 & 14325 & 0.0013 & & 77 & 1825 & 0.042 & & 96 & 16150 & 0.0059\\
\hline
          & 69 & 28351 & 0.0024 & & 210 & 3106 & 0.068 & & 279 & 31457 & 0.0089\\
\hline
\label{tab:allstars}
\end{tabular}
\tablenotetext{\dag}{Categorization by the $P$ (stellar rotation period estimated from lightcurve). $P>10d$ (Slow), $P<10d$ (Fast) and the total of them (Total).} 
\tablenotetext{\ddag}{The number of flare stars ($N_{\rm fstar}$), the number of all observed G-type dwarfs ($N_{\rm all}$), and the fraction of 
stars that flare ($f_{\rm fstar}=N_{\rm fstar}/N_{\rm all}$) in the category.}
\end{center}
\end{table}

\begin{table}
   \begin{center}
      \caption{Parameters of superflare Sun-like stars}
      \begin{tabular}{ccccccc}
      \hline
      KeplerID & $\teff$ \tablenotemark{\dag}(K)& $\logg$ \tablenotemark{\ddag} & $R/R_{\odot}$ \tablenotemark{\dag\dag}& KPmag&$P_{\rm rot}(day)$ \tablenotemark{\star} & \# of flares\\
      \hline

        5522535 & 5732 & 4.3 & 1.3 & 13.8 & 20.3 & 1\\
        6750902 & 5654 & 4.4 & 1.1 & 14.7 & - & 1\\
        6865484 & 5688 & 4.4 & 1.1 & 13.8 & 11.2 & 12\\
        7133671 & 5657 & 4.4 & 1.1 & 15.5 & 15.8 & 1\\
        7354508 & 5714 & 4.4 & 1.1 & 13.4 & 17.0 & 1\\
        7597685 & 5834 & 4.6 & 0.9 & 15.9 & 21.8 & 1\\
        8212826 & 5811 & 4.2 & 1.4 & 14.0 & 26.3 & 2\\
        8880526 & 5936 & 4.3 & 1.2 & 12.8 & - & 1\\
        9574994 & 5925 & 4.4 & 1.1 & 15.1 & 12.3 & 1\\
        9766237 & 5674 & 4.6 & 0.9 & 13.9 & 21.8 & 1\\
        9944137 & 5725 & 4.6 & 0.8 & 13.8 & 25.3 & 1\\
        10471412 & 5771 & 4.1 & 1.6 & 13.4 & 15.1 & 4\\
        10524994 & 5747 & 4.5 & 1.0 & 15.3 & 12.0 & 5\\
        11390058 & 5785 & 4.3 & 1.3 & 12.6 & 12.1 & 3\\
        11401109 & 5732 & 4.5 & 0.9 & 14.5 & 29.1 & 1\\
        11455711 & 5664 & 4.7 & 0.8 & 14.0 & 13.9 & 3\\
        11494048 & 5929 & 4.4 & 1.1 & 13.4 & 14.9 & 1\\
        11612371 & 5826 & 4.4 & 1.1 & 13.4 & - & 2\\
        11961324 & 5750 & 4.4 & 1.1 & 14.2 & - & 2\\

    \end{tabular}
    \tablenotetext{\dag}{Effective temperature}
    \tablenotetext{\ddag}{Surface gravity}
    \tablenotetext{\dag\dag}{Stellar radius in unit of the solar radius}
    \tablenotetext{\star}{Stellar rotation period estimated from brightness variation}
    \label{tab:sunlike}
  \end{center}
\end{table}

\end{document}